\documentstyle[12pt]{article}
\title{Universal geometric approach to uncertainty,
entropy and information}
\author{Michael J. W. Hall\\ \\Abteilung f\"{u}r Quantenphysik\\
Universit\"{a}t Ulm\\D-89069 Ulm, Germany\\ \\Address after September
30 1988:\\Theoretical Physics, IAS\\Australian National University\\
Canberra ACT 0200, Australia}
\date{}
\begin{document}
\maketitle
\newpage
\begin{abstract}
It is shown that  a unique measure of {\it volume} is associated
with any statistical ensemble, which directly quantifies the 
inherent spread or localisation of the ensemble.  It is
applicable whether the ensemble is classical or quantum, 
continuous or discrete, and may be derived from a small number of
theory-independent geometric postulates.  Remarkably, this unique
{\it ensemble volume} is proportional to the exponential of the 
ensemble entropy, and hence provides a novel geometric characterisation
of the latter quantity.  Applications include unified 
volume-based derivations of the Holevo and Shannon bounds 
in quantum and classical information theory, a precise
geometric interpretation of thermodynamic entropy for equilibrium
ensembles, a geometric derivation of semi-classical uncertainty
relations, a new means for defining classical and quantum localization
for arbitrary evolution processes, a geometric interpretation of
relative entropy,
and a new proposed definition for the spot-size of 
an optical beam.  Advantages of ensemble volume over other measures
of localization (root-mean-square deviation, Renyi entropies,
and inverse participation ratio) are discussed.

PACS Numbers: 03.65.Bz, 03.67.-a, 05.45+b, 42.60.Jf
\end{abstract}

\renewcommand{\thesection}{\Roman{section}}
\renewcommand{\thesubsection}{\Alph{subsection}}

\section{INTRODUCTION}

This paper has two main goals.  The first is to demonstrate that 
for any ensemble, whether classical, quantum, discrete or continuous,
there is essentially only {\it one} measure of the ``volume''
occupied by the ensemble which is compatible with basic geometric
notions.  This {\it ensemble volume} is thus a preferred and 
universal choice for characterising what is variously referred to as
the spread, dispersion, uncertainty, or localisation of an ensemble.

Remarkably, the derived ``ensemble volume'' turns
out to be proportional to the
exponential of the entropy of the ensemble.  A by-product of the first
goal is thus a new universal characterisation of ensemble entropy, based
on geometric notions.   Indeed, a number of
properties of ensemble entropy turn out to have simple geometric 
interpretations.  The universal nature of the characterisation is of
particular interest:  the only previous {\it context-independent} 
interpretation
of ensemble entropy to date (and hence
applicable in particular to ensembles
described by continuous probability distributions) 
appears to be as a somewhat vague measure of
uncertainty or randomness.  
 
The second goal is to apply ``ensemble volume'' to a wide range of
contexts in which ensembles appear.  The applications demonstrate
not only the advantages of ensemble volume over other measures
 of spread, but also to some extent why it is that  
ensemble entropy makes a natural appearance in
contexts as diverse as statistical mechanics, information theory, chaos,
and quantum uncertainty relations. Some results have been briefly
reported elsewhere \cite{eprint}.  Here important details and
extensions are given, as well as a number of new results. 

The work reported here was originally motivated by several connections 
between volume and information.  Shannon proved an
upper bound on information transfer, via classical signals subject to
quadratic energy and noise constraints, by considering ratios of 
spherical volumes in high-dimensional spaces \cite{shannon}.  
One can similarly obtain 
approximate upper bounds on information
for {\it quantum} signals, via semi-classical arguments
involving ratios of phase space volumes \cite{cd,hallpra},
which in some cases turn out to be exact.
This raises the question
of whether there is some general  measure of volume which
can be used to derive rigorous information bounds for the general
case.  This question is answered affirmatively here, and a new unified
derivation of the classical Shannon 
and the quantum Holevo information bounds is given,
based on simple volume properties.

There are also a number of connections which have been made
previously between volume and entropy.  For example, derivations in
statistical mechanics 
typically obtain heuristic expressions for thermodynamic entropy by 
counting ``microstates'' in a phase-space 
volume of ``small'' thickness containing a
constant-energy surface \cite{statmech}.
Ma in an interesting approach attempted to
{\it define} the thermodynamic entropy of a system in classical
statistical mechanics as proportional to 
the logarithm of a phase-space volume
corresponding to the ``region of motion'' of the system \cite{ma}, 
although
he could not rigorously define the latter region.  A {\it precise
geometric} interpretation of thermodynamic entropy for both classical
and quantum equilibrium ensembles will be given here.

Further, Leipnik introduced the exponential of the position entropy of
a quantum system as a measure of its ``volume'', and favourably
compared the 
associated uncertainty relations for position and momentum with the
usual Heisenberg uncertainty relations \cite{leipnik} (see also the
review in
\cite{manko} and Sec. II.C below).  Generalisations to other measures
of ``volume'' were given by Zakai \cite{manko,zakai}. It is demonstrated
here that the former measure has a unique geometrical
significance, and a geometrical derivation of quantum
uncertainty relations is given based on the property that quantum
states have a minimum ensemble volume.

Zyczkowski \cite{zed} and more recently Mirbach and Korsch \cite{mk1,mk2}
have used entropy as a
measure of  ``localisation'' for   
chaotic quantum and classical systems 
for various initial states.
The results of the present paper show that this measure can be simply
related to the spread of ensemble volume for arbitrary evolution
processes, and provide support for the use of
this measure over all other localisation
measures.

Rather than going immediately to general postulates for volume, 
and formal proofs of uniqueness,
the following section first explores 
ensemble volume 
for a familiar class of ensembles:  those described by 
one-dimensional probability distributions.  In this case 
the ensemble volume reduces to a ``length'', 
which is calculated for a number of concrete
examples and compared with other measures of uncertainty such as
root-mean-square deviation.  Geometric
properties of this ``length'' and an associated quantum 
uncertainty relation are discussed.
Two-dimensional joint  probability distributions are also briefly
discussed, where the ensemble volume becomes an
``area'' that is geometrically related 
to the ``lengths'' of the marginal distributions.  This ``area''
motivates a new definition for the spot size of an optical beam.

In Section III and an accompanying appendix, the derivation
of the ensemble volume from universal geometric postulates is given. 
These postulates depend on theory-independent notions of invariance,
projection onto orthogonal axes, and additivity, and in particular are
independent of whether the ensemble is classical or quantum.  
The bonus of a new geometrical characteristion of ensemble entropy is
discussed, and a  geometrical interpretation of relative entropy is 
given.

Applications to statistical mechanics, semi-classical quantum mechanics,
information theory, chaos and other types of dynamical evolution 
are given in Section IV. Conclusions are presented in Section V.

\section{1- AND 2-DIMENSIONAL EXAMPLES}

Before deriving the unique form of ensemble volume in
Sec.III, it is useful to first consider some of its properties and
connections to other measures of uncertainty in two 
familiar settings:  continuous distributions on the line and on 
the plane,
for which ``volume'' reduces to the special cases of 
``length'' and to ``area'' respectively.
These special cases are already 
sufficient to exemplify a number of general features of
ensemble volume, and its advantages as a measure of spread. 

\subsection{Length}

Consider a 1-dimensional probability distribution $p(x)$, corresponding
to some random variable $X$ (e.g., position, momentum, or phase).  
There are then a number of candidates for 
a direct measure of 
the ``uncertainty'' or ``spread'' of $X$, the
most well known being the root-mean-square (RMS) deviation
\begin{equation} \label{rms}
\Delta X = [\int dx \, x^2 p(x) - (\int dx \, x p(x))^2 ]^{1/2}  .
\end{equation}
This quantity is a ``direct'' measure in the sense of 
having the same units as $X$, and has the virtues of being
invariant under translations and reflections, scaling linearly with
$X$ ($\Delta Y =$ $\lambda \Delta X$ for $Y=\lambda X$),
and vanishing in the 
limit that $X$ has some definite value $x'$. 

A second candidate is the inverse participation ratio \cite{zed,
mk2,hell}
\begin{equation} \label{part}
\xi_{X} = [\int  dx \, p(x)^2 ]^{-1} ,
\end{equation}
(which may also be recognised as a monotonic function of the so-called
``linear entropy'' $-\int dx\, p(x)^2$ \cite{zurek}).  
This quantity shares all of the above-noted virtues
of $\Delta X$.  However, it is in fact only
a special case of what may be called the ``Renyi length''
\begin{equation} \label{renyi}
L_{X,\alpha} = [\int dx \, p(x)^{1+\alpha} ]^{-1/\alpha} \hspace{2cm} 
(\alpha\geq -1) 
\end{equation}
(named for its logarithm - a generalised entropy defined by Renyi
\cite{renyi}). Renyi lengths are directly related to measures of
uncertainty considered by Zakai for quantum systems \cite{manko,zakai},
and use of their reciprocals
as (indirect) measures of uncertainty 
has been extensively investigated in \cite{thesis} (see also 
\cite{maas}).  The inverse participation ratio corresponds to $\alpha=1$
in Eq.~(\ref{renyi}).

The Renyi length $L_{X,\alpha}$ in Eq.~(\ref{renyi}) 
satisfies all of the above-noted properties 
of $\Delta X$ (same units as $X$, translation/reflection invariance,
scaling linearly with $X$, and vanishing as $p(x)$ approaches a delta
function).  
Eq.~(\ref{renyi}) thus introduces an uncountable infinity of possible
candidates for a direct measure of uncertainty! Fortunately, as will be
seen in Sec.III, just one of these Renyi lengths may be singled out 
uniquely over all other possible measures on geometric grounds.

In particular, in this paper special attention will be paid to the case 
$\alpha \rightarrow 0$ in Eq.~(\ref{renyi}).  
The corresponding length
will simply be denoted by $L_{X}$, and is just the exponential
of the usual ensemble entropy \cite{foot1}:
\begin{equation} \label{length}
L_{X} = L_{X,0} = \exp [-\int dx \, p(x) \ln p(x)]  .
\end{equation}
It is a special case of the ``ensemble volume'' to be derived
in Sec.III, and will therefore be referred to as the 
{\it ensemble length}.  

\subsection{Comparisons}

In Table I the RMS deviation and ensemble length are calculated for
several types of 1-dimensional distributions. As noted following
Eqs.~(\ref{rms}) and (\ref{renyi}) both quantities are invariant under
translations and scale linearly with $X$.  Hence they can be trivially
calculated for distributions of the form $p(x/a - x')/a$ once they
have been found for $p(x)$ (by simply multiplying the result for the 
latter case by $a$). The Table will be used to highlight
a number of differences between $\Delta{X}$ and $L_{X}$. 

First, it is seen from Table I that the ensemble length exists in
cases when the RMS deviation does not (for Cauchy-Lorentz and 
sink-squared distributions in particular). It may further be shown that
$L_{X}$ is finite whenever $\Delta{X}$ is: 
the well known variational property that ensemble entropy is maximised
for a fixed value of $\Delta{X}$ by a Gaussian distribution
\cite{ash} immediately implies from the scaling property and Table I
that 
\begin{equation} \label{rmsineq}
L_{X} \leq (2\pi e)^{1/2} \Delta{X}  .
\end{equation}
Thus the use of ensemble length as a measure of
uncertainty allows a wider quantitative range of applicability than
does RMS deviation.  
This permits, for example, the quantitative discussion
of quantum uncertainty relations, expressed in terms of ensemble
length, for cases in which the usual Heisenberg
uncertainty relations have nothing to say (see following subsection).

Second, the calculations for the uniform and circular
distributions, $p_{U}$ and $p_{C}$ in Table I respectively, 
exemplify a maximality property of ensemble length: it
is maximised on a given interval by a uniform
distribution on the interval, with a 
maximum value equal to the length of
the interval.  Thus one may write
\begin{equation}
L_{X} \leq L
\end{equation}
for a distribution confined to an interval of length $L$ \cite{foot2}.
This property 
reflects the intuitive notion that that $p(x)$ is most spread out or
least localised when it is {\it flat}, having no peaks where 
probability is concentrated. The RMS deviation does not conform to this
notion, achieving its maximum possible value in the limit of two
maximally-separated peaks (a distribution equally concentrated on the 
endpoints of the interval).

Third, the calculation in Table I for the uniform 
and double-uniform
distributions $p_{U}$ and $p_{DU}$ illustrates an addivity property 
of ensemble length: the ensemble length of
$p_{DU}$ is twice that of the two non-overlapping 
uniform distributions $p_{U}(x-a)$ and 
$p_{U}(-x-a)$ which it comprises in equal mixture. More generally,
if $p(x)$ and $q(x)$ denote two non-overlapping distributions of
equal ensemble length $L$, then any mixture $\lambda p(x)$ $+$
$(1-\lambda)q(x)$ of these distributions satisfies
\begin{equation} 
L_{X} \leq 2L ,
\end{equation}
with the upper bound achieved for $\lambda =1/2$ \cite{foot3}. 
This property
reflects the intuitive notions that such a mixture is least localised
(most spread out) when it is not more concentrated in one of the
non-overlapping regions than in the other, and that 
for this equally-weighted case the non-overlapping lengths simply add.
In contrast, the RMS deviation of
$p_{DU}$ depends strongly on the separation of the peaks, and 
indeed becomes infinite as this separation increases.  This
example and the one above emphasise what 
can be directly seen from Eq.~(\ref{rms}):
the RMS deviation is a measure of {\it separation} of the 
region(s) of concentration from a particular point of the distribution
(the mean value), rather
than a measure of the extent to which the distribution is in
fact concentrated.

Fourth, except in cases where 
the second moment of $p(x)$ has some particular
physical meaning, it is difficult to assess the significance
of a given value of $\Delta X$ without some further information
about the distribution.  For example, even for single-peaked
distributions, the probability that $X$ lies within
$\pm \Delta X$ of the mean is highly dependent upon the nature
of $p(x)$ \cite{foot4}. 
In contrast, as will be
seen in Sec. III, the ensemble length $L_{X}$ has a unique
geometrical significance.
 
Finally, it is of interest to make a 
quantitative comparison between the 
degrees to which a given distribution $p(x)$ is concentrated in a
region of length $L_{X}$ on the one hand, and of length $2\Delta X$
on the other.  To do so, it is natural to define
the {\it maximum confidence} corresponding to a given length $L$ as
\begin{equation}
C(L) = \sup_{\{A:\mid A\mid=L\}} \{ \int_A dx\, p(x) \}  ,
\end{equation}
where the supremum is over all measurable sets $A$ of total length $L$.
In the case of a distribution symmetric about a single peak this is
achieved by choosing $A$ to be the interval of length $L$ centred
on the mean value of the distribution.

From Table I one can calculate the values of $C(L_{X})$ to be
approximately 100\%, 99\%, 96\%, 93\%, 91\% and 90\% for the
uniform, circular, gaussian, exponential, sink-squared and 
Cauchy-Lorentz distributions respectively.  The corresponding values of 
$C(2\Delta X)$ are 58\%, 61\%, 68\%, 86\% for the first four of the
above distributions, with the value being undefined for the last two.
It is seen that for these examples  $C(L_{X})$ varies over a much
narrower range than $C(2\Delta X)$, and that $L_{X}$ typically 
corresponds to a larger confidence value than $2\Delta X$.

\subsection{Uncertainty relations}

The relationship between ensemble length and ensemble entropy in 
Eq.~(\ref{length}) allows the usual entropic uncertainty relation
for the position and momentum of a quantum particle \cite{bbm}
to be equivalently written in the geometric form
\begin{equation} \label{uncert}
L_{X} L_{P} \geq \pi e\hbar , 
\end{equation}
relating the product of the ensemble lengths to a minimum area in
phase space.
Bounding $L_{X}$ and  $L_{P}$ from above via 
Eq.~(\ref{rmsineq})  then immediately yields the well known 
Heisenberg uncertainty relation
\begin{equation} \label{heis}
 \Delta X \Delta P \geq \hbar /2  .
\end{equation}

The above two inequalities are similar in form, and have the
same broad physical significance: the particle cannot
be prepared in a state for which both the position and
momentum distributions have arbitrarily small spreads. However, it 
is seen that the latter inequality is mathematically weaker, 
as it follows from the former. For example, it follows 
from Eq.~(\ref{uncert}) that $L_{P}$ (and hence, via 
Eq.~(\ref{rmsineq}), $\Delta P$) becomes
infinite as $p(x)$ approaches a weighted sum of delta functions.  This
cannot be concluded from Eq.~(\ref{heis}). 

Inequality (\ref{uncert}) may used to make quantitative
evaluations regarding the relative spreads of position and momentum
in cases where the Heisenberg inequality (\ref{heis}) yields {\it no}
information.  For example, consider a quantum particle confined to
an interval of length $L$, such that the position amplitude is constant
over the interval.  It follows that the momentum statistics are 
described by  the sink-squared distribution
\begin{equation} \label{sink}
\pi^{-1} (2\hbar /L) (\sin [pL/(2\hbar)]/p)^2   .
\end{equation}
As noted in Table I the RMS deviation
$\Delta P$ is not defined in this case, and hence the Heisenberg 
inequality cannot be used
to assess  the degree to which  position and momentum are jointly
localised. In contrast, using Eq.~(\ref{sink}), Table I and the 
scaling property of ensemble length, one finds 
\begin{equation} \label{sinkuncert}
L_{X} L_{P} = 2\pi\exp[2(1-C)]\hbar \approx 15\hbar ,
\end{equation}
where $C\approx 0.577 215 66$ denotes Euler's constant.
Hence the particle has an associated phase space area close to the
lower bound of $\pi e\hbar\approx 9\hbar$ in Eq.~(\ref{uncert}); i.e.,
the particle is in fact in an approximate minimum
uncertainty state of position and momentum.  

A similar example
is the case of a particle confined to the positive $x$-axis, with 
a position amplitude that decays exponentially with $x$.  
The position and momentum distributions are then given by exponential
and Cauchy-Lorentz distributions of the forms $p_{E}(x/a)/a$ and
$2ap_{CL}(2ap/\hbar)/\hbar$ respectively, implying via Table I and
the scaling property that
\begin{equation}
 L_{X} L_{P} = 2\pi e \hbar . 
\end{equation}
Hence the state is relatively 
well-localised in position and momentum, with an associated
phase-space area only twice that of the minimum in Eq.~(\ref{uncert}).
Again, the  Heisenberg
uncertainty relation Eq.~(\ref{heis}) gives no information about the 
joint localisation in this case. 

Finally, it may be mentioned that
there is an uncertainty relation relating the Renyi lengths of position
and momentum for general $\alpha$:  it follows from Eq.~(131) of 
\cite{manko} that
\begin{equation} \label{renineq}
L_{X,\alpha} L_{P,\beta} \geq \pi\hbar [1+2\alpha]^{1+1/(2\alpha)}
/(1+\alpha)
\end{equation}
for $\alpha\geq -1/2$, where $\beta =-\alpha /(1+2\alpha )$.
For $\alpha=\beta=0$ the lower bound is maximum, and
the inequality reduces to Eq.~(\ref{uncert}) above.

\subsection{Area and spot size}

This section will be concluded by briefly looking at measures of spread 
for {\it two}-dimensional distributions, to highlight a further
geometric property of ensemble length of importance 
in later sections.  This property also holds for RMS deviation,
but not for Renyi lengths in general.  A related measure of 
spot size for optical beams is defined and briefly discussed.

Each of the ``length'' measures in Eqs.~(\ref{rms}), (\ref{renyi}) and
(\ref{length}) has a natural generalisation to a measure of ``area'',
corresponding to the spread or uncertainty
of a 2-dimensional probability distribution
$p(x,y)$ of two random variables $X$ and $Y$:
\begin{eqnarray} \label{rmsarea}
\Delta A & = & [\det (\langle {\bf xx^{T}}\rangle - 
\langle{\bf x}\rangle \langle{\bf x^{T}}\rangle)]^{1/2}  , 
\\ \label{renarea}
A_{XY,\alpha} & = & \langle p^\alpha \rangle^{-1/\alpha}  ,
\\ \label{area}
A_{XY} & = & \exp[\langle -\ln p \rangle ]
\end{eqnarray}
respectively, where ${\bf x}$ denotes the column vector $(x,y)$,
${\bf x^T}$ its transpose, and $\langle\cdot\rangle$ the average
with respect to $p$.  These areas satisfy properties analogous to
to their 1-dimensional counterparts, and will be referred to as the
RMS area, Renyi area, and ensemble area respectively.

The RMS area in Eq.~(\ref{rmsarea}) may be recognised as the product of
the RMS deviations along the principal axes of the distribution in the
$xy$-plane, and in general
satisfies the inequality (Eq.~(2.13.7) of \cite{hardy})
\begin{equation} \label{rmsin}
\Delta A \leq \Delta X \Delta Y ,
\end{equation}
with equality for the case that $p(x,y)$ factorises into two
uncorrelated distributions for $X$ and $Y$.

This inequality for ``area'' and ``length'' has a simple geometric
interpretation, to be generalised in the following Section.  
In particular, 
the marginal distributions $p_{1}(x)$ and $p_{2}(y)$ for $X$ and $Y$
are obtained by "projecting"  the joint distribution $p(x,y)$
onto the two orthogonal $x$ and $y$ axes. The associated RMS lengths
$\Delta X$ and $\Delta Y$ may be similarly thought of as obtained by
``projecting'' the RMS area $\Delta A$ onto these axes.  However, this
is only consistent with Euclidean geometry if inequality 
(\ref{rmsin}) holds: the product of the two lengths obtained by
projection of an area onto two orthogonal axes can never be less than
the original area.

Ensemble area and ensemble length are also consistent with this 
``projection'' interpretation:  the well known
subadditivity of entropy \cite{ash} can be equivalently 
written via Eqs.~(\ref{length})  and (\ref{area}) as 
\begin{equation} \label{areaproj}
A_{XY} \leq L_{X} L_{Y} , 
\end{equation}
in analogy to Eq.~(\ref{rmsin}). The subadditivity of entropy is thus
seen to correspond to a projection property of Euclidean geometry.  
One has the further related property that
if $p(x,y)$ is uniform on a rectangular region oriented parallel to
the $x$ and $y$ axes, and vanishes outside this region, then equality
holds in Eq.~(\ref{areaproj}) with $L_{X}$ and $L_{Y}$ corresponding
to the lengths of the sides of the rectangle.  
Thus Eq.~(\ref{areaproj}) reduces in this case to 
the Euclidean property {\it area $=$ length $\times$ breadth}.
In general, the Renyi areas in Eq.~(\ref{renarea}) are not consistent 
with the projection property, as will be seen in Sec.~III.  

Finally, it may be noted that Eq.~(\ref{area}) may be applied to 
physical distributions other than probability distributions,
with corresponding geometrical advantages.
For example, let $P(x,y)$ denote the time-averaged power distribution
in some plane orthogonal to the direction of propagation of an optical
beam.  One may then define the ``geometric''  spot size of the beam as
the ensemble area of the normalised power distribution
$P(x,y)/P_{T}$, where $P_{T}$ is the integrated power over the plane:
\begin{equation}
A_{geom} = P_{T} \exp [-(P_{T})^{-1} \int dxdy \,
P(x,y) \ln P(x,y)]  .
\end{equation}
This satisfies desirable properties such as being additive for
non-overlapping identical beams, being invariant with respect to
scaling the power up or down, scaling linearly with 
beam magnification, having a maximum value of $A$ for a beam
confined to an area $A$ (attained for a uniform power distribution
over that area), and satisfying a ``projection property''
analogous to Eq.~(\ref{areaproj}).  It is also invariant under any
transformation of coordinates which preserves area in the usual sense 
(i.e., with
unit Jacobian), and so to this extent is independent of the 
coordinatisation of the plane.
Alternative definitions based on, for example,
Eqs.~(\ref{rmsarea}) or (\ref{renarea}) are geometrically less 
satisfying.

\section{ENSEMBLE VOLUME}

The previous section indicates the wide range of possible measures for 
the spread of one- and two-dimensional probability distributions, and
draws attention to a number of geometric and other advantages
enjoyed by the ``length'' and ``area'' defined in 
Eqs.~(\ref{length}) and (\ref{area}) respectively.

As noted in the Introduction, it has 
often proved useful to employ various notions of ``volume''
for statistical ensembles across a wide variety of contexts,
such as information theory, statistical mechanics, uncertainty 
relations, and chaotic  evolution.  Other contexts 
include Ornstein-Uhlenbeck diffusion and 
semi-classical quantum mechanics (see \cite{eprint} and Secs. IV.B and
IV.D below). 
This raises the question of whether there is  in
fact some {\it universal} 
measure of ``volume'' for classical and quantum
ensembles, which may be usefully employed in all of the above contexts
and which is not restricted in application or interpretation to 
various special cases.

Here it will be shown that indeed such a measure exists, 
which may be uniquely derived from a small number of theory-independent
postulates fundamental to the concept of ``volume''.  It generalises
the ensemble length and ensemble area of the previous section, and will 
be referred to as the {\it ensemble volume}.  It also leads to new
geometric characterisations of entropy and relative entropy.

\subsection{Notation}

Three generic types of ensemble will be considered here. 
The first is a classical
ensemble described by a continuous probability distribution 
$p({\bf x})$ on some $n$-dimensional space $X$; the
second is a classical ensemble described by a discrete probability
distribution $\{ p_{i}\}$ where $i$ ranges over some discrete set 
$I$; and the third is a quantum ensemble described by a
density operator $W$ on some Hilbert space $H$.  

Each of the above types of ensemble shares some universal features.
It is essential to abstract a number of these features via a common
notation if ``volume'' is to be discussed in a theory-independent
manner. 

For example, consider the three identities
\begin{equation} \nonumber
\int_{X} d^n{\bf x}\, p({\bf x})=1,\,\sum_{i\in I} p_i =1,\, 
{\rm tr}_{H}[W]=1 .
\end{equation}
Defining $\Gamma$ to correspond respectively to the spaces/sets 
$X$, $I$ and $H$;
${\rm Tr}_{\Gamma}[\cdot]$ to correspond 
respectively to integration over $X$,
summation over $I$, and the trace over $H$; and $\rho$ to correspond
respectively to the ensembles $p({\bf x})$, $\{ p_{i}\}$, and 
$W$; these identities
can be subsumed into the generic identity 
\begin{equation} \label{norm}
{\rm Tr}_{\Gamma}[\rho] = 1 . 
\end{equation}

Another universal feature is the notion of {\it composite} or
{\it joint} ensembles: for a given pair of spaces/sets $\Gamma_1$,
$\Gamma_2$ of a given type one can define a composite set/space
$\Gamma_{12}$, where for classical and quantum ensembles 
$\Gamma_{12}$ corresponds to the set product and the tensor
product respectively of $\Gamma_1$ and $\Gamma_2$. 
Further, if 
$\rho$ describes a composite ensemble on $\Gamma_{12}$ one may
define two {\it projected} ensembles $\rho_1$, $\rho_2$ on $\Gamma_1$
and $\Gamma_2$ respectively, via
\begin{equation} \label{marg}
\rho_1 = {\rm Tr}_{\Gamma_2}[\rho],\hspace{1cm}
\rho_2= {\rm Tr}_{\Gamma_1}[\rho] .
\end{equation}
These projected ensembles correspond to {\it marginal}
distributions and {\it reduced} density operators for the cases of
classical and quantum ensembles respectively.
 
Finally, one may define any two ensembles $\rho$, $\rho'$ of 
the same type to be {non-overlapping} if and only if
\begin{equation}
{\rm Tr}_{\Gamma}[\rho\rho'] = 0 .
\end{equation}
Note that in general two ensembles are non-overlapping if and only if 
they can be distinguished by measurement without error.

\subsection{Postulates for volume}

For the three types of ensemble discussed in the previous subsection
it is useful to think of ``volume'' in the following ways.  First, for
a continuous distribution $p({\bf x})$ 
on a space $X$, the volume corresponds
to a direct measure of the region of ``spread'' of $p({\bf x})$ in $X$.
Second, for a classical discrete distribution $\{ p_i\}$, 
one may imagine
the indices as labelling a set of boxes or bins. 
In this case ``volume'' corresponds to the spread of  
the distribution over these bins, i.e., as a continuous 
measure of the effective number of bins occupied by the distribution.
Third, for a quantum ensemble, the volume may be considered as a 
continous generalisation of Hilbert space dimension, corresponding to
a measure of the spread of the ensemble in Hilbert space. 

Consider now a measure of volume, $V(\rho)$, which satisfies the
following properties:

{\it (i) Invariance Property:}  $V(\rho)$ is invariant under all 
transformations on $\Gamma$ which preserve ${\rm Tr}_{\Gamma}[\cdot]$ 
(these are represented by measure-preserving transformations on $X$ for 
continous classical ensembles, permutations on $I$ for discrete
classical ensembles, and unitary transformations on $H$
for quantum ensembles). 

{\it (ii) Cartesian Property:}  If $\rho$ describes two {\it
uncorrelated}
ensembles $\rho_{1}$ and $\rho_{2}$ on $\Gamma_{1}$ and $\Gamma_{2}$
respectively, then 
\begin{equation} \label{cart}
V(\rho) = V(\rho_{1}) V(\rho_{2}) 
\end{equation}
(note $\rho$ is the product $\rho_{1}\rho_{2}$ for classical ensembles,
and
the tensor product $\rho_{1}$$\otimes$$\rho_{2}$ for quantum ensembles).

{\it (iii) Projection Property:}  If $\rho$ describes an ensemble of
composite systems on $\Gamma_{12}$ then
\begin{equation} \label{proj}
V(\rho) \leq V(\rho_{1}) V(\rho_{2}) ,
\end{equation}
where $\rho_{1}$, $\rho_{2}$ are the projections of $\rho$ 
defined in Eq.~(\ref{marg}).

{\it (iv) Additivity Property:}  An equally-weighted mixture of $m$
{\it non-overlapping} ensembles $\rho_a$, $\rho_b$, $\dots$,
each of equal  volume $V$, has a total volume of $mV$, i.e.,
\begin{equation} \label{add}
V(m^{-1}[\rho_a + \rho_b + \dots]) = m V .
\end{equation}

{\it (v) Uniformity Property:}  If $\rho$ is any mixture of $m$
non-overlapping ensembles of equal volumes $V$, then
\begin{equation} \label{unif}
V(\rho) \leq m V .
\end{equation}

The above properties are essentially the same as those defined in 
\cite{eprint}, where the additivity and uniformity properties were
combined in the latter.  Their geometrical significance 
is as follows.

First, the invariance property (i) 
ensures that the volume $V(\rho)$ is a
function of the ensemble alone, independently of a particular 
co-ordinatisation, labelling, or measurement basis for $\Gamma$.
Indeed, the transformations which preserve ${\rm Tr}_{\Gamma}[\cdot]$
are exactly those which preserve volume, or measure, on $\Gamma$
in the usual sense.
For example, for a classical distribution $p({\bf x})$
on $X$ the measure of a subset $S\subseteq X$ is given by 
\begin{equation}
\mid S\mid = \int_S d^n{\bf x} = Tr_{S}[1] .
\end{equation}
The invariance property then requires that the ensemble volume is
invariant under all transformations which preserve the measure of
all subsets, i.e.,
those transformations with unit Jacobian.  For the case of a
classical phase space such transformations include all 
canonical transformations, and hence $V(\rho)$ will be
invariant under Hamiltonian evolution. One may similarly consider
the measure $\mid S\mid = {\rm Tr}_{S}[1]$ of 
subsets $S\subseteq I$ and subspaces
$S\subseteq H$; in these cases the invariance property again requires 
that $V(\rho)$ is invariant under measure-preserving transformations,
corresponding to permutations and unitary transformations respectively. 

Second, the Cartesian property (ii) is exactly analogous to the 
geometric property that area equals length times breadth, and more
generally that the volume of the Cartesian product of two sets is
equal to the product of the volume of the sets.  This is illustrated
in Figure 1.  

Third, the projection property (iii) is exactly analogous to the 
geometric property that a volume is less than or equal to the product
of the lengths obtained by its projection onto orthogonal axes, and is
illustrated in Figure 2.  It is a generalisation of the projection
property discussed for RMS area and ensemble area in Sec. II.D.

Fourth, the additivity property (iv) requires the ensemble volume to
be additive for a uniform mixture of 
non-overlapping ensembles of equal volume.  The geometric interpretation
of this is self-evident: the total volume of 
$m$ equal non-overlapping volumes is the sum of the individual volumes.

Finally, the uniformity property (v) states that the maximum volume, of
a mixture of non-overlapping ensembles of 
equal volume, is bounded by the
sum of the component volumes.  Thus, noting the additivity
property, this maximum is achieved for a {\it uniform} 
mixture, i.e., one which is not more concentrated
on one of the component ensembles than on any other.

\subsection{Derivation}

Here the unique, universal measure of volume for ensembles is obtained.
It may more generally be applied as a measure of spread for
any positive classical or
quantum density, such as beam intensity or mass density, by
calculating the ``volume'' of the corresponding normalised density.
In such cases, where no ensemble is involved, one could alternatively 
label this quantity as the ``geometric dispersion''.

In particular, one has the following result, first stated in 
\cite{eprint}, and proved in the Appendix:

{\it Theorem:}  Any (continuous) measure of volume 
satisfying properties (i)-(v)
above has the form
\begin{equation} \label{theo}
V(\rho) = K(\Gamma) e^{S(\rho)},
\end{equation}
where $S(\rho)$ denotes the ensemble entropy
\begin{equation} \label{ent}
S(\rho) = - {\rm Tr}_{\Gamma}[\rho \ln \rho] ,
\end{equation}
and $K({\Gamma})$ is a constant which may depend on $\Gamma$, and 
satisfies
\begin{equation} \label{kgam}
K(\Gamma_{12}) = K(\Gamma_{1}) K(\Gamma_{2}) .
\end{equation}

The proof in the Appendix primarily relies on 
applying properties (i)-(v) to an arbitrarily large number
of independent copies of a given ensemble $\rho$.  I believe it may
be possible to prove the theorem without the uniformity 
property (v), but have not been able to do so.

The constant $K(\Gamma)$ in Eq.~(\ref{theo}) is a 
normalisation constant, reflecting the notion that only
relative volumes are of real interest in comparing different ensembles.
For continuous classical ensembles a natural choice is $K(\Gamma)=1$,
so that a distribution which is uniform over a set $S$ of measure $V$,
and vanishes outside $S$, has ensemble volume equal to $V$.  

For discrete classical ensembles the choice $K(\Gamma) = 1$ 
corresponds to measuring the 
ensemble volume in terms of the number of ``bins'' occupied by the
ensemble,  with the minimum volume of 1 bin corresponding to a
distribution with $p_i = 1$ for some index $i$.
However, if the distribution arises from the
discretisation of a continuous observable such as position (due to
measurement limitations for example), then it would be natural to
choose $K(\Gamma)$ to correspond to the discretisation volume.
If the index set is finite, with $M$ labels, another
possible choice for $K(\Gamma)$ is $1/M$.  
The ensemble volume then
measures the fraction of the total volume 
occupied by the ensemble.

For quantum ensembles the choice $K(\Gamma) = 1$ 
corresponds to measuring the
ensemble volume in terms of the number of Hilbert space dimensions
occupied by the ensemble, with pure states occupying the minimum
possible of 1 dimension.  However, if the Hilbert space $H$ has finite
dimension $M$ then one could alternatively take $K(\Gamma) = 1/M$,
corresponding to a fractional measure of volume in analogy to the
classical case.  Finally, 
for quantum systems with classical counterparts, 
such as spin-zero particles, one may choose $K(\Gamma)$ so that
in the classical limit the quantum ensemble volume reduces to the
classical ensemble volume.  This is explored further in Sec. IV.B, and
used to obtain semi-classical uncertainty relations. 

It should be noted that the assumption of continuity in the statement
of the theorem is necessary.  For example, one may for a discrete 
classical ensemble $\{ p_i\}$ define the ``support volume'' as the
number of non-zero $p_i$ values.  This satisfies all of properties
(i)-(v), but
is not continuous. The simplest counterexample is the 
discrete probability distribution  $\{1-\epsilon,
\epsilon\}$ for $\epsilon >0$.  As $\epsilon
\rightarrow 0$ this distribution continuously approaches the 
distribution $\{ 1, 0\}$, with a support volume of 1; 
however for all $\epsilon >0$ the support volume is 2. 
  
If one defines the ``RMS'' volume for an $n$-dimensional observable
${\bf x}$ by generalising Eq.~(\ref{rmsarea}) to arbitrary dimensions
\cite{footrms}, it is not difficult to show 
that the invariance property
(restricted to {\it linear} transformations), 
the Cartesian property, and the
projection property are satisfied.  However
it does not satisfy the additivity and 
uniformity properties.
Further, the ``Renyi'' volumes
\begin{equation} \label{renvol}
V_{\alpha}(\rho ) = ({\rm Tr}_{\Gamma}[\rho^{1+\alpha}])^{-1/\alpha} ,
\end{equation}
defined in analogy with the Renyi length and Renyi area in Eqs.
(\ref{renyi}) and (\ref{renarea}) respectively, satisfy properties (i),
(ii), (iv) and (v) for all $\alpha\geq -1$.  However, a counterexample
given by Renyi (Theorem 4 of Sec. IX.6 in \cite{renyi}) shows that
the projection property is {\it not} satisfied, except for the cases
$\alpha =0$ (corresponding to Eq.~(\ref{theo}) with $K(\Gamma)=1$),
and $\alpha =-1$ (corresponding to the discontinuous case of ``support
volume'' discussed above).

\subsection{Geometric characterisation of entropy}

The appearance of the ensemble entropy in Eq.~(\ref{theo}) as a result
of geometric postulates (i)-(v) provides a new approach to this 
quantity, which is moreover independent of whether the ensemble is
classical or quantum, discrete or continuous.
In particular, {\it ensemble entropy may be defined} (up to an 
additive constant) {\it as the logarithm of the ensemble volume}, 
where the
latter is taken to be the primary quantity.  The properties of ensemble
entropy may thus be regarded as being geometric in origin. 
Indeed, it will be 
seen that its natural appearance in a number of physical contexts
can be interpreted as following from its relationship to a ``volume''.

The geometric interpretation of ensemble entropy contrasts markedly 
with its only other 
context-independent interpretation as an (indirect) measure of
``uncertainty'' or ``randomness'' \cite{renyi,thesis,maas,ash,
shan,foot5}.  Indeed ensemble volume provides a {\it direct} measure of 
uncertainty, which is advantageous when one wishes to compare the
spreads of two
ensembles of a given type (i.e., with the same $\Gamma$).  For example, 
if two ensembles have
entropies of 0.5 bits and 1.5 bits respectively \cite{footbit}, should 
one compare
their ratio or their difference in assessing the degree to which
the uncertainty of the first exceeds that of the second? Since entropies
are typically only defined up to a multiplicative constant 
(see below), one might consider the ratio to be the more 
signicant means of comparison.  However, the ensemble
volume gives an unequivocal answer: the volume of the second ensemble
is twice that of the first in this case, and hence has twice
the spread. 

It is interesting to briefly compare the derivation of ensemble volume 
from properties (i)-(v) with existing axiomatic derivations of ensemble 
entropy. Such axiomatic derivations are reviewed in \cite{behara}, and
are all related to the original derivation given by Shannon 
\cite{ash,shan}.  Unlike the theorem of the previous section
they are limited to {\it discrete} classical ensembles. Moreover, 
they lead to an arbitrary multiplicative constant for entropy,
whereas the geometric approach leads to an arbitrary 
{\it additive} constant for entropy.

To see that the axioms used by Shannon and others 
are markedly different from 
properties (i)-(v) used to derive ensemble volume, consider the
``grouping axiom'' of Shannon \cite{shan} 
(see also Sec. 1.2 of \cite{ash}), which may
be written in the notation of this paper as:
\begin{equation} \label{shanax}
S(\lambda \rho + (1-\lambda)\rho') = S(\{\lambda, 1-\lambda\})
+ \lambda S(\rho) + (1-\lambda) S(\rho ')
\end{equation}
for any two non-overlapping discrete classical ensembles $\rho$, 
$\rho '$.  Thus it is assumed that the ``randomness'' $S(\cdot )$ of a
mixture of non-overlapping distributions is equal to that of
the mixing distribution plus the average randomness of the 
individual ensembles. This axiom, together with a continuity assumption
and a symmetry assumption equivalent to the invariance property (i),
is sufficient to derive the form $S(\rho ) = -C \sum_i p_i \ln p_i$ for 
the entropy of discrete classical ensembles, where $C$ is an arbitrary
constant \cite{behara}.

Eq.~(\ref{shanax}) does {\it not} translate 
into a natural axiom for ensemble
volume: replacing $S$ by $\ln V$ gives the equivalent
constraint
\begin{equation}
V(\lambda \rho + (1-\lambda)\rho') = V(\{\lambda, 1-\lambda\})
[V(\rho)]^\lambda [V(\rho')]^{1-\lambda} ,
\end{equation}
which has no simple geometric interpretation.  Conversely, the 
additivity property Eq.~(\ref{add}), that non-overlapping equal
volumes add, translates under $V\rightarrow \exp S$ 
into the ``randomness'' constraint
\begin{equation}
S(\rho /2 + \rho '/2) = \ln 2 + S ,
\end{equation}
which is not a natural property to postulate 
for a measure of ``randomness''.
The geometric approach to ensemble entropy given here thus differs
significantly from former approaches (as is also apparent from comparing
the proof in the Appendix with those in \cite{ash,shan,behara}).

Finally, it is of interest to note that the concavity property of
ensemble entropy, $S(\sum_i \lambda_i \rho_i) \geq \sum_i \lambda_i
S(\rho_i)$ \cite{ash,shan}, is equivalent to an inequality relating
the volume of a mixture to the weighted geometric mean of
the volumes of its components:
\begin{equation} \label{mixture}
V(\sum_i \lambda_i \rho_i) \geq \prod_i [V(\rho_i)]^{\lambda_i}  .
\end{equation}
This may be regarded as a generalisation of the uniformity property
Eq.~(\ref{unif}), as it implies that uniform mixtures have the 
greatest volumes.  
Note that the ensemble volume may
itself be regarded as a weighted geometric mean (e.g., of the function
$p(x)^{-1}$ with respect to $p({\bf x})$ for continous classical 
ensembles; see sections 2.2 and 6.7 of \cite{hardy}).
 
\subsection{Relative entropy}

The relative entropy of two ensembles $\rho$ and $\sigma$ may be
defined in a context independent manner by \cite{relent}
\begin{equation} \label{relent}
S(\rho\mid\sigma) =  {\rm Tr}_\Gamma [\rho (\ln \rho - \ln \sigma)] .
\end{equation}
It is asymptotically related to the probability 
of mistaking ensemble $\rho$ for 
ensemble $\sigma$, as is reviewed in \cite{plenio}.
Here it will briefly be indicated how a geometric interpretation of this
quantity can be given.

Consider a compact $n$-dimensional space $X$ which is divided up into 
into a set of non-overlapping bins $\{ B_i\}$ (e.g., for measurement
purposes).  A discrete probability distribution $\{ p_i\}$ over the bins
(e.g., corresponding to measurement results), may then also be
modelled by the {\it continuous} distribution $p({\bf x})$ on $X$ 
defined by
\begin{equation} \label{cont}
p({\bf x}) = p_i /V_i , \hspace{1cm} {\bf x}\in B_i ,
\end{equation}
where $V_i = \int_{B_i} d^n{\bf x}$ denotes the measure of bin $B_i$.  
Thus $p({\bf x})$ is uniform over each bin, 
and its integral over bin $B_i$ is
equal to $p_i$. Let $\rho_D$ and $\rho_C$ denote the discrete and
continuous ensembles 
corresponding to $\{ p_i\}$ and $p({\bf x})$ respectively.

Now, as discussed earlier, the ensemble volume $V(\rho_D )$ is
proportional to the effective number of bins occupied by $\rho_D$.
However, this does not indicate the effective volume or spread of
the ensemble relative to $X$, particularly in the case of varying
bin-sizes $V_i$.  The latter is given by $V(\rho_C)$, which, making the
choice $K(\Gamma)=1$, follows from Eq.~(\ref{cont}) as
\begin{equation} \label{rhoc}
V(\rho_C ) = \exp [-\sum_i p_i \ln (p_i /V_i ) ] .
\end{equation}
Note that in the case of {\it equal} 
bin-sizes $V_i \equiv V$ this reduces to
the bin-size $V$ multiplied by the effective number of bins occupied,
$\exp S(\rho_D)$. 

Finally, if $X$ has total measure $\sum_i V_i = V_X$, one may define
the ``weighting'' ensemble $\sigma_D$ as corresponding to the 
discrete probability distribution $\{ V_i /V_X\}$.  Thus $\sigma_D$
describes the relative sizes or weightings of the bins.  
It then follows via
Eqs.~(\ref{relent}) and (\ref{rhoc}) that
\begin{equation} \label{ratio}
V(\rho_C)/V_X = e^{-S(\rho_D\mid\sigma_D)} .
\end{equation}

Hence {\it the relative entropy} $S(\rho\mid\sigma)$ {\it is directly 
related to 
the volume of a discrete ensemble}
 $\rho$ {\it embedded in a continuous space,
where} $\sigma$ {\it characterises the distribution of bin sizes of the
embedding}.
Note that this geometric interpretation of relative entropy allows its 
properties to be understood as corresponding to ratios of volumes.  For
example, the volume of an ensemble on $X$ can never be greater than
$V_X$ (corresponding to a uniform distribution on $X$). Hence
the left-hand-side of Eq.~(\ref{ratio}) is never greater
than unity, implying that
\begin{equation}
S(\rho\mid\sigma) \geq 0 .
\end{equation}

\section{APPLICATIONS}

The results of Sec. II for ensemble length and ensemble area indicate
the usefulness of ensemble volume as a direct measure of the spread of
an ensemble (and of other positive densities such as optical beam 
power).  Here other applications will be
examined, in the contexts of statistical mechanics, semi-classical 
quantum mechanics, information theory, and quantum chaos.  
A particular
result of note is a new unified proof of the classical Shannon
information bound and the quantum Holevo information bound based on
ratios of ensemble volumes.  For the quantum case this proof is 
conceptually and technically far simpler than previous proofs.

\subsection{Statistical mechanics}

First, in the statistical mechanics context, 
the Gibbs relation $S_{th}=k S(\rho)$ between thermodynamic entropy and
ensemble entropy for equilibrium ensembles can be rewritten via 
Eq. (\ref{theo}) as
\begin{equation} \label{gibb}
S_{th} = k \ln [V(\rho)/K(\Gamma)] .
\end{equation}
Thus, the thermodyamic entropy is 
(up to an additive constant) 
proportional to the logarithm of the ensemble volume.  

From  Eq.~(\ref{gibb}) and the third law of thermodynamics 
(that thermodynamic entropy vanishes at absolute zero), it follows that
one should choose $K(\Gamma)$ to correspond to 
a minimum ``zero-temperature" ensemble volume.  For quantum
ensembles one has from  Eqs. (\ref{theo}) and (\ref{ent}) that
$V(\rho)=K(\Gamma)$ for pure states, i.e., the 
{\it quantum} zero-temperature volume is just that of a 
{\it pure} state on $\Gamma$.  Similarly, for discrete classical
ensembles, $K(\Gamma)$ is the volume of the ``pure'' ensemble 
described by $\{ 1, 0, 0, \dots\}$. 
However, continuous classical ensembles violate the
third law \cite{statmech} and  $K(\Gamma)$ remains arbitrary in this
case (but see Sec. IV.B below). 

The geometric expression (\ref{gibb}) is very similar 
to the original Boltzmann relation
\begin{equation} \label{bolt}
S_{th} = k \ln W ,
\end{equation}
where $W$ is the number of distinct
microstates or ``elementary complexions" consistent with the
thermodynamic description.  Indeed, 
from the above discussion it follows that
Eq.~(\ref{gibb}) provides a {\it precise} 
{\it geometric} interpretation
of the Boltzmann relation for discrete classical and quantum
equilibrium ensembles:  {\it thermodynamic 
entropy is proportional to the logarithm of  
the number of non-overlapping zero-temperature 
volumes contained
within the total volume of the ensemble}.  Thus the Boltzmann
relation and the Gibbs formula for thermodynamic  entropy become
{\it directly}
unified in the ensemble volume approach, without appeals to reservoirs,
microcanonical ensembles, etc.

Properties of thermodynamic entropy can be reinterpreted in terms
of geometric volume.  For example, the additivity of thermodynamic
entropy for uncorrelated ensembles in thermal equilibrium follows from
Eq.~(\ref{gibb}) and the Cartesian property Eq.~(\ref{cart}) for 
uncorrelated ensemble volumes. Note also
that irreversible processes correspond geometrically to those 
which increase the volume of the ensemble.

\subsection{Semi-classical quantum mechanics}

Consider now a classical ensemble $\rho_{C}$ which is the
``classical limit" of some quantum ensemble $\rho_{Q}$, i.e., the
physical properties of $\rho_{C}$ approximate those of
$\rho_{Q}$.  Such ensembles exist, for example, for
equilibrium ensembles in the high-temperature limit and for the
coherent states of a harmonic oscillator. 

For the case of a spinless particle associated with a 
$2n$-dimensional phase space one can obtain a relationship between the
constants $K(\Gamma_{C})$ and $K(\Gamma_{Q})$ in Eq.~(\ref{theo}) by
requiring that the ensemble volumes $V(\rho_{C})$ and $V(\rho_{Q})$
are approximately equal for such ensembles.  Since these constants are
independent of the dynamics of the ensemble it suffices to choose
an equilibrium ensemble of
isotropic oscillators. Equating the calculated values of
$V(\rho_{C})$ and $V(\rho_{Q})$ in the high-temperature
limit then yields
\begin{equation} \label{corr}
K(\Gamma_{Q}) = h^{n}  K(\Gamma_{C})  ,
\end{equation}
for the volume of a pure state, where $h$ is Planck's constant.
Thus the Bohr-Sommerfeld quantization rule that a pure quantum
state occupies a phase-space volume of $h^n$ is recovered
\cite{semiclass}. 

Eq.~(\ref{corr}) can be used to derive semi-classical uncertainty
relations from geometric considerations.  For two
corresponding ensembles  $\rho_{Q}$ and  $\rho_{C}$ as above
the position and momentum entropies $S_X$ and $S_P$ respectively
must be approximately equivalent for either ensemble. Further, 
\begin{equation} \label{projxp}
\exp(S_X) \exp(S_P) \geq \exp(S(\rho_{C}))
\end{equation}
holds for the classical ensemble from the projection property 
Eq.~(\ref{proj})
applied to projections onto the position and momentum axes. 
Eqs. (\ref{theo}), (\ref{corr}) and
(\ref{projxp}) then yield the approximate inequality
\begin{equation} \label{entrop}
S_X + S_P - S(\rho_Q) \stackrel{>}{\sim} n \ln h  
\end{equation}
for quantum ensembles which have classical limits.
I conjecture that {\it exact} inequality in fact 
holds for {\it all} quantum
ensembles.

Since the entropy of a quantum ensemble has a minimum value of 0 
(corresponding to the existence of a minimum volume for quantum 
ensembles), it follows from Eq.~(\ref{entrop}) that one has the 
semi-classical entropic uncertainty relation
\begin{equation} \label{enta}
S_X + S_P \stackrel{>}{\sim} n \ln h ,
\end{equation}
for quantum ensembles with classical limits.
As per the derivation of Eq.~(\ref{heis}) from 
Eq.~(\ref{uncert}), the
corresponding semi-classical Heisenberg uncertainty relation
\begin{equation} \label{heisa}
\Delta X \Delta P \stackrel{>}{\sim} \hbar/e
\end{equation}
then follows for the $n=1$ case.  Eqs. (\ref{enta}) and (\ref{heisa}) 
are close to the exact results for
general quantum ensembles \cite{manko,bbm} (see Eqs.~(\ref{uncert})
and (\ref{heis})).
It is seen that geometrically
they correspond to application of the projection property
Eq.~(\ref{proj}) to the
projections of a pure state of volume $h^n$ onto the position and 
momentum axes (i.e., replacing $\Gamma_1$ and $\Gamma_2$ by
$X$ and $P$ in Figure 2).

\subsection{Information bounds}

Consider a communication channel
where signals represented by ensembles $\rho_{1}$, $\rho_{2}$, $\dots$ 
are transmitted with
prior probabilities $p_{1}$, $p_{2}$, $\dots$ respectively
\cite{footsig}.  
The ensemble of signal
states itself corresponds to the mixture
\begin{equation} \label{ensem}
\rho = \sum_{i} p_{i} \rho_{i} .
\end{equation}
For classical ensembles it was shown by Shannon \cite{ash,shan}
that the average amount of error-free data $I$ which can be obtained per
transmitted signal, measured in terms of the number of binary digits
required to represent the data, is bounded above by
\begin{equation} \label{shanhol}
I \leq [S(\rho) - \sum_i p_i S(\rho_i)]/\ln 2  .
\end{equation}
The formally equivalent bound for quantum ensembles was proved by Holevo
\cite{holevo}, and hence Eq.~(\ref{shanhol}) may be referred to as
the Shannon-Holevo information bound.

Proofs given in the literature of Eq.~(\ref{shanhol}) for the quantum 
case are mathematically rather technical in nature, and quite different
in character to proofs for the classical case 
\cite{holevo,others}.  However, the formal equivalence of the
quantum and classical bounds suggests that a unified proof exploiting
universal features of statistical ensembles may be possible.  Indeed the
construction of such a proof, based on simple volume arguments, 
was recently outlined in \cite{eprint}, and will be elaborated on here. 
A second such proof, which reduces the general classical/quantum case 
to that
of discrete classical noiseless channels, will also be pointed out.

First, consider a message consisting of $L$ signals chosen from the
set $\{\rho_i\}$.  Such a message may be denoted by $\rho_{\alpha}$,
where $\alpha = (i_1 , i_2 , \dots , i_L )$ denotes the labels of the
signals comprising the message. In the limit that $L\rightarrow\infty$
the strong law of large numbers implies that the relative frequency of
signal $\rho_i$ appearing in the message approaches $p_i$ with
probability 1.  It follows from the Cartesian property Eq.~(\ref{cart})
that the volume of the message satisfies
\begin{equation} 
V(\rho_\alpha ) \rightarrow V_{mess} = \prod_i [V(\rho_i )]^{p_i L} ,
\end{equation}
as $L\rightarrow\infty$.  Moreover, as will be shown below in 
Eq.~(\ref{rhol}), the volume
of any ensemble of such messages is bounded above by $[V(\rho)]^L$.
Hence, using the additivity property Eq.~(\ref{add}),  the maximum 
possible number of non-overlapping messages of length $L$, $N_L$, 
satisfies 
\begin{equation} \label{rat} 
N_L \leq [V(\rho)]^L /V_{mess} 
\end{equation}
as $L\rightarrow\infty$.  Noting that {\it error-free} data can
only be obtained from distinguishing 
among a set of {\it non-overlapping} messages, and that $N_L$ 
such messages require at most $1 + \log_2 N_L$ binary digits to record,
it follows in the limit of infinitely long messages that the
average information gained per signal, $I$, is bounded by
\begin{equation} \label{made}
I \leq \lim_{L\rightarrow\infty} L^{-1} (1 + \log_2 N_L) \leq
\log_2  V(\rho)/\prod_i [V(\rho_i )]^{p_i}  .
\end{equation}
Finally, since communication based on finite message lengths cannot
transmit more data per signal than communication based on infinite
lengths, the bound holds for all signalling schemes, and 
Eq.~(\ref{shanhol}) follows from Eqs.~(\ref{theo}) and (\ref{made}).

The above proof of the Shannon-Holevo bound is geometrically simple,
being based on the ratio of the maximum available volume for an
ensemble of messages to the message volume (Eq.~(\ref{rat})). 
Note that the argument cannot be used to derive similar bounds based on
other invariant volume measures, as all of the defining properties of
ensemble volume are required. However, heuristic arguments of the
same type for other volume measures can sometimes give excellent results
\cite{cd,hallpra}. Note  that the
Shannon-Holevo bound is in fact {\it tight} for both classical and
quantum ensembles \cite{ash,shan,tight}, corresponding geometrically to
being able to choose a number $N_L$ of messages 
arbitrarily close to the upper bound in Eq.~(\ref{rat}) which can be 
distinguished with a vanishingly small average error probability as
$L\rightarrow\infty$.  

To conclude this subsection it will be shown that the Shannon-Holevo
bound may also be proved by considering only messages of finite length,
and applying the classical noiseless coding theorem
\cite{ash, shan}.  With notation as above, suppose that one chooses a
set of codewords $C$ from the set of messages of length $L$, and that 
codeword $\rho_\alpha \in C$ is transmitted with probability 
$q(\alpha)$.  Defining $N_{i}(\alpha)$ as the number of times signal
$\rho_i$ appears in codeword $\rho_\alpha$, and $\overline{\rho}_l =
\sum_{\alpha\in C} q(\alpha) \rho_{i_l}$ as the average $l$-th 
component of the transmitted codewords, consistency requires that
\begin{eqnarray} \label{consist}
p_i & = & L^{-1} \sum_{\alpha\in C} q(\alpha) N_i (\alpha) ,\nonumber\\
\rho & = & L^{-1} \sum_{l=1}^{L} \overline{\rho}_l  .
\end{eqnarray}
Using the projection property Eq.~(\ref{proj}) and Eq.~(\ref{mixture}) 
one then has the  inequality chain
\begin{equation} \label{rhol}
V(\sum_\alpha q(\alpha)\rho_\alpha ) \leq V(\overline{\rho}_{1})\dots 
V(\overline{\rho}_{L})
\leq [V(\sum_l L^{-1}\overline{\rho_l})]^{L} = [V(\rho )]^L  . 
\end{equation}
To obtain a bound for error-free data, it must be assumed that the 
codewords are non-overlapping, so that they can be distinguished without
error by measurement.  From Eq.~(\ref{theo}) and the Cartesian property
Eq.~(\ref{cart}) one may then calculate
\begin{equation}
V(\sum_\alpha q(\alpha)\rho_\alpha ) = e^{S[q]} \prod_{\alpha\in C}
[V(\rho_\alpha)]^{q(\alpha)} = e^{S[q]} \prod_{\alpha\in C}
\prod_l [V(\rho_{i_l})]^{q(\alpha)} ,\nonumber 
\end{equation}
where $S[q]$ denotes the entropy of the discrete distribution
$\{q(\alpha)\}$.  Combining this with Eqs.~(\ref{consist}) and 
(\ref{rhol}) then gives
\begin{eqnarray}
S[q] & \leq & L S(\rho) - \sum_{\alpha\in C} \sum_l
q(\alpha) S(\rho_{i_l}) \nonumber \\
& = & L S(\rho) - \sum_{\alpha\in C} \sum_i q(\alpha) N_i (\alpha)
S(\rho_i ) \nonumber \\
& = & L [ S(\rho) - \sum_i p_i S(\rho_i ) ]  .
\end{eqnarray}
Finally, from Shannon's classical noiseless coding theorem \cite{ash,
shan} $S[q]/\ln 2$ is the maximum information (measured in binary
digits) which can be transmitted on average per codeword, and hence 
Eq.~({\ref{shanhol}) follows for the average information transmitted per
signal.

\subsection{Chaotic and other diffusion processes}

Zyczkowski \cite{zed} and Mirbach and Korsch \cite{mk1,mk2} have 
studied connections between quantum and classical chaos via 
entropies associated with the evolution of coherent states.  Here it
will be shown that this approach may be simply interpreted in terms of
ensemble volume, and considerably generalised.  

Consider an ensemble $\rho_0$, classical or quantum, which evolves in 
time under some dynamical process $D$ (not necessarily reversible).  The
ensemble will explore some region of $\Gamma$, which may
be large for standard diffusion processes, or relatively small for
integrable and dissipative systems.  The localisation of the ensemble in
$\Gamma$ over time is characterised by the time-averaged mixture
\begin{equation} \label{time}
\overline{\rho} = \lim_{T\rightarrow\infty} T^{-1} \int_{0}^{T} dt\,
\rho_t .
\end{equation}
This mixture gives greatest weight to regions of $\Gamma$ where the
ensemble spends the most time. Hence its ensemble volume,
$V(\overline{\rho})$, is a
measure of the spread of the region explored by the ensemble as it
evolves.  

The {\it localisation ratio} for a given initial state and dynamical
process
may now be defined as the ratio of the volumes of $\overline{\rho}$
and $\rho_0$, i.e.,
\begin{equation} \label{rati}
r = V(\overline{\rho})/V(\rho_0) = \exp [S(\overline{\rho}) -
S(\rho_0) ] .
\end{equation}
It thus measures the localisation of the ensemble under the
evolution process, relative to
its initial spread.  This ratio will be less than or equal to one if
the ensemble evolves to a fixed point, and greater than or equal to
one if it diffuses over the whole of $\Gamma$.  For chaotic systems with
integrable regions it will depend strongly on the initial ensemble.
The above definition is clearly 
natural on geometric grounds, and the ensemble entropy
appears as a consequence of the uniqueness theorem in Eq.~(\ref{theo}).

For classical and quantum systems corresponding to the same 
evolution process, it is of interest to compare localisation properties.
This is easily done for the case of initial quantum ensembles
$\rho_Q$ which have corresponding classical counterparts $\rho_C$  (such
as coherent states).  In this case the quantum and classical 
localisation ratios $r_Q$ and $r_C$ can be calculated and compared.
Zyczkowksi partially carries through this procedure in \cite{zed}, 
where he plots $S(\overline{\rho})$ 
for the quantum counterpart of a classically chaotic process,
where $\rho_Q$ is chosen to range over a set of coherent 
states indexed by their corresponding phase-space points.  In this case 
$S(\overline{\rho})$ is just the entropy of the
energy distribution of $\rho_Q$. Noting 
$S(\rho_Q)=0$ for pure states, it
follows from Eq.~(\ref{rati}) that this is equivalent to plotting the
logarithm of the localisation ratio, $\ln r$.  However, he compares
quantum localisation features qualitatively with the classical phase 
space portrait,
rather than quantitatively with analogously calculated classical 
localisation ratios.

Mirbach and Korsch extended the approach of Zyczkowski by also 
calculating
$S(\overline{\rho})$ for the classical ensembles $\rho_C$ corresponding
to the coherent states $\rho_Q$.  For a complete family of such states
they then compared the corresponding classical and quantum values of 
$S(\overline{\rho})$ (Figures 1 and 3 of \cite{mk2}).
Since for this case $S(\rho_Q)$ and $S(\rho_C)$ are constants, this
amounts to comparing the logarithms of the classical and quantum
localisation ratios (up to an additive constant).

However, Mirbach and Korsch argue that one should in fact compare 
{\it measurement} entropies rather than the direct ensemble entropies,
to smear out quantum fluctuations in the latter case \cite{mk1,mk2}.
This is also easily interpreted in terms of localisation ratios.
In particular, for a measurement observable $A$ on a classical or 
quantum ensemble $\rho$, let
$V_{A}(\rho)$ denote the volume of the measurement distribution of $A$.
The localisation ratio of an evolution process 
with respect to $A$, for an initial ensemble $\rho_0$, is then defined
in analogy to Eq.~(\ref{rati}) as
\begin{equation}
r_A = V_A (\overline{\rho})/V_A (\rho_0 )  .
\end{equation} 
Again one may compare localisation ratios for classical and quantum
ensembles, where one chooses corresponding observables $A_Q$ and $A_C$.
The logarithm of this quantity (up to an additive constant) is plotted
in Figures 2 and 3 of \cite{mk1} for quantum and classical systems
respectively for a complete set of coherent states, 
where $A_C$ is chosen to be a phase-space measurement 
(so that $r_{A_C} = r_C$), and $A_Q$ to be
a ``Husimi'' phase-space measurement corresponding to the
complete set of coherent states \cite{husimi}.

\section{CONCLUSIONS}

In conclusion, an essentially unique measure of volume for classical
and quantum ensembles has been found, related to ensemble entropy,
which provides a 
geometric tool for any context in which ensembles appear.
This measure is  universal in the sense that it may
be defined by theory-independent concepts of invariance, 
uncorrelated ensembles, projection, and non-overlapping ensembles
(properties (i)-(v)).  

Its properties as a direct measure of ``spread''
have been investigated in Sec. II for continuous distributions, and
favourably compared with measures based on root-mean-square deviation.  
New geometric characterisations of ensemble entropy and relative entropy
have been discussed in Secs. III.D and III.E.

Applications include a new definition of spot size for optical beams;
a precise geometric
interpretation of the Boltzmann relation in statistical mechanics; 
a derivation of semi-classical uncertainty relations based on 
the existence of a minimum volume for quantum states and a
projection property of volumes; a unified
derivation of results in classical and quantum information theory
based on simple volume ratios; 
and a new and universal definition of a localisation ratio which
measures the time-averaged spreading of an ensemble and underlies
entropic measures previously 
investigated in the context of quantum chaos.

Work is in progress on further applications, particularly 
to quantum information theory \cite{tight}, measures of quantum
entanglement \cite{plenio}, and information exclusion relations
\cite{hallpra, hallprl}. The conjecture suggested following
Eq.~(\ref{entrop}) is also under active investigation, and the (mostly 
weaker) bound
\begin{equation}
S_X + S_P - S(\rho) \geq \ln 2\pi e\hbar - 
\ln [1 + \Delta X \Delta P/(\hbar/2)]
\end{equation}
has thus far been found for the $n=1$ case.

{\bf Acknowledgments}

I am grateful to Prof. Wolfgang Schleich for drawing my attention
to the inverse participation ratio 
(thus stimulating the search for the 
``best'' measure of volume), and to Prof. Hajo Leschke and 
Drs. Gernot Alber and Bruno Mirbach for useful discussions.
This work was supported by the Alexander von Humboldt Foundation.


{\bf APPENDIX}

Here the fundamental 
theorem stated in Sec. III.C is proved, showing essentially 
that the exponential of the ensemble entropy is the unique measure
of the volume of a statistical ensemble.  It is convenient to first
prove the theorem for discrete classical ensembles, and then extend
the arguments to quantum ensembles and to continuous classical 
ensembles.  Notation will be as defined in Sec. III.A, and reference
will be made to the five assumed properties of the volume measure
$V(\rho)$ stated in Sec. III.B.

Let $\rho$ denote a classical discrete ensemble $\{ p_i\}$, with finite 
index set $I = \{ 1,2, \dots ,M\}$.  Defining the ``pure'' ensemble
$\rho_j$ ($j\in I$)  as corresponding to the distribution 
$\{p^{(j)}_{i}\}$ with $p^{(j)}_{i} = \delta_{ij}$, one can write 
$\rho$ as the mixture
\begin{equation} \label{rhomix}
\rho = \sum_{i\in I} p_i \rho_i  .
\end{equation}
Note that one has the two basic properties
\begin{equation} \label{pure}
{\rm Tr}_{\Gamma}[\rho_j \rho_k] = 0 \,\, (j\not= k),\hspace{1cm}
V(\rho_j ) = {\rm constant} = V_I . 
\end{equation}
The first states that these pure ensembles are non-overlapping, and
the second that they have equal ensemble volumes (this follows from the 
invariance property, noting that the $\rho_j$
map to each other under permutations).

Now consider the ensemble $\rho^L\in\Gamma^L$ corresponding to $L$ 
uncorrelated copies of $\rho$.  For
each $\alpha = (i_1, i_2, \dots , i_L)$ in $I^L$ define
\begin{equation} \label{rhoalpha}
\rho_{\alpha} = \rho_{i_1}\rho_{i_2}\dots\rho_{i_L} , 
\hspace{1cm} p(\alpha) = p_{i_1}  p_{i_2} \dots  p_{i_L} .
\end{equation}
Thus $\rho_{\alpha}$ corresponds to the uncorrelated composite ensemble 
formed by $\rho_{i_1}, \rho_{i_2}, \dots , \rho_{i_L}$ (in that order).
One can then decompose $\rho^L$ into the mixture
\begin{equation} \label{mix}
\rho^L = \sum_{\alpha\in I^L} p(\alpha) \rho_{\alpha}  .
\end{equation}
The proof of the theorem proceeds by finding a suitable set of
so-called ``typical sequences'' $T\subseteq I^L$ 
\cite{ash,shan}, which allows $\rho^L$
in Eq.~(\ref{mix}) to be approximated by certain mixtures
of the ensembles $\{\rho_{\alpha}\}$ where $\alpha$ is restricted to
range over $T$.

For a given $\alpha\in I^L$ let $N_i(\alpha)$ denote the number of
times the index $i$ appears as a 
component of  $\alpha$, and let $P(\alpha)\in I^L$ correspond to a
permutation of the components of $\alpha$.  If $S(\rho)$
denotes the entropy of $\rho$ defined in Eq.~(\ref{ent}) of the
text, then for any
$\epsilon > 0$ and $L$ sufficiently large 
one may choose a set $T$, with $\mid T\mid$ elements, which satisfies:
\begin{eqnarray}
& (T1) & C_T = \sum_{\alpha\in T} p(\alpha) > 1-\epsilon , \nonumber\\
& (T2) & \mid T\mid = e^{L[S(\rho)+\delta_L]} , \nonumber\\
& (T3) & \sum_{i\in I} \mid L^{-1} N_{i}(\alpha) - p_i\mid < \delta_{L}'
\,\, {\rm for \,\, all}\,\, \alpha \in T ,  \nonumber \\
& (T4) & \alpha\in T \,\,{\rm implies}\,\, P(\alpha)\in T \,\,
{\rm for\,\, all}\,\, P  ,
\nonumber
\end{eqnarray}
where both $\delta_L$ and $\delta_{L}' \rightarrow 0$ as 
$L\rightarrow\infty$.
A particular example of such a set is 
\begin{equation} \label{Tdef}
T = \{ \alpha :\, \mid L^{-1}N_{i}(\alpha) - p_i\mid <
[Mp_i (1-p_i )/(L\epsilon )]^{1/2} \} .
\end{equation}
Properties (T1) and (T2) for this set
are proved in Theorem 1.3.1 of \cite{ash};
property (T3) follows noting that $\sum_i [p_i (1-p_i )]^{1/2}$ is
bounded by $(M-1)^{1/2}$ and hence that one can choose
$\delta_{L}' = M(L\epsilon)^{-1/2}$; and
property (T4) is an immediate consequence of 
$N_{i}(\alpha)$ being invariant under permutations.

To obtain an upper bound for the volume $V(\rho)$ of $\rho$, 
consider now the ensembles defined by the mixtures
\begin{equation}  \label{tmix}
\rho_L (T) = C_T^{-1}
\sum_{\alpha\in T} p(\alpha) \rho_\alpha , \hspace{1cm}
\rho_{L}^{*}(T) = \mid T\mid^{-1} \sum_{\alpha\in T} \rho_\alpha , 
\end{equation}
where $C_T = \sum_{\alpha\in T} p(\alpha)$.
From the Cartesian property and Eqs.~(\ref{pure}) and (\ref{rhoalpha})
it follows that $V(\rho_\alpha )=[V_I]^L$ is constant, and further that
the $\rho_\alpha$ are non-overlapping.  Hence, from the uniformity
and additivity properties, $V(\rho_L (T)) \leq V(\rho_{L}^{*}(T)) =
\mid T\mid [V_I]^L$.  Property (T2) then gives
\begin{equation} \label{ding}
V(\rho_L (T)) \leq [V_I]^L e^{L[S(\rho)+\delta_L]}  .
\end{equation}
Further, from property (T1) and Eqs.~(\ref{mix}) and (\ref{tmix}),
\begin{eqnarray}
{\rm Tr}_{\Gamma^L}[\mid\rho^L - \rho_L (T)\mid ] & = & 
\sum_{\alpha\in T}
\mid p(\alpha) - p(\alpha)/C_T \mid + \sum_{\alpha\notin T} p(\alpha)
\nonumber\\ & = & (1/C_T - 1) C_T + (1 - C_T ) \leq 2\epsilon\nonumber .
\end{eqnarray}
Hence $\rho^L$ can be made arbitrarily close to $\rho_L (T)$ for $L$
sufficiently large, and so from the assumed continuity of $V(\cdot)$,
and noting from the Cartesian property that $V(\rho^L) = [V(\rho)]^L$,
one has from Eq.~(\ref{ding}) that 
\begin{equation} \label{less}
V(\rho) = \lim_{L\rightarrow\infty} [V(\rho_L (T))]^{1/L} \leq 
V_I e^{S(\rho)}  .
\end{equation}
Thus the exponential of the entropy is an upper bound for the ratio of
the volume of $\rho$ to the volume of a ``pure'' state.
Note that only properties (T1) and (T2) of $T$ were needed to obtain 
this result, and that the projection property has not been used.

To obtain the converse of inequality Eq.~(\ref{less}), note from the
projection property that 
\begin{equation} \label{wow}
V(\rho_{L}^* (T)) \leq \prod_{l=1}^{L} V(\overline{\rho}_l (T)) ,
\end{equation}
where $\overline{\rho}_l (T)$ is the projection of $\rho_{L}^* (T)$ onto 
its $l$-th component, i.e.,
\begin{equation} \label{sevtwo}
\overline{\rho}_l (T) = \sum_{\alpha = (i_1 , \dots i_L ) \in T} 
|T|^{-1} \rho_{i_l}  .
\end{equation}
From property (T4) of $T$, $\overline{\rho}_l (T)$ is
independent of $l$  and hence may be denoted by $\overline{\rho}$.
Eq.~(\ref{wow}) then becomes 
$V(\rho_{L}^* (T)) \leq [V(\overline{\rho})]^L$. 
But as noted earlier, 
the volume of $V(\rho_{L}^* (T))$ follows from the additivity property
as $\mid T\mid [V_I]^L$, and hence via property (T2) of $T$ 
Eq.~(\ref{wow}) reduces to
\begin{equation} \label{sing}
V_I e^{S(\rho) + \delta_L} \leq V(\overline{\rho})  .
\end{equation}
Further, from Eqs.~(\ref{tmix}) and (\ref{sevtwo}) 
\begin{equation}
\overline{\rho}=L^{-1}\sum_l \overline{\rho}_l (T) = \mid T\mid^{-1} 
\sum_{\alpha\in T} \sum_{i\in I} L^{-1} N_{i}(\alpha) \rho_i  ,
\end{equation}
and hence from Eq.~(\ref{rhomix}) and property (T3) of $T$ 
\begin{eqnarray}
{\rm Tr}_{\Gamma}[\mid\rho - \overline{\rho}\mid] & = & \mid T\mid^{-1}
{\rm Tr}_{\Gamma}[\mid \sum_{\alpha\in T} \sum_{i\in I} (p_i -  
L^{-1} N_{i}(\alpha)) \rho_i \mid ] \nonumber\\
  & \leq & \mid T\mid^{-1} \sum_{\alpha\in T} \sum_{i\in I}
\mid p_i -L^{-1} N_{i}(\alpha) \mid \leq \delta_{L}' \nonumber  .
\end{eqnarray}
Hence $\overline{\rho}$ can be 
made arbitrarily close to $\rho$ for $L$ sufficiently
large, and so, taking the limit $L\rightarrow\infty$ in 
Eq.~(\ref{sing}), the assumed continuity of $V(\cdot)$ gives 
\begin{equation} \label{more}
V_I e^{S(\rho)} \leq V(\rho)  .
\end{equation}

Eqs.~(\ref{less}) and (\ref{more}) yield the theorem of Sec. III.B
for classical discrete ensembles with finite index sets (where
$K(\Gamma)$ in Eq.~(\ref{theo}) is identified with the volume $V_I$ of
a pure ensemble $\{ p_i = \delta_{ij}\}$ on $I$, and Eq.~(\ref{kgam})
for $K(\Gamma)$ follows immediately from the Cartesian property).  
The extension to ensembles with
infinite index sets is trivial by continuity. The distribution
$\{ p_i\}$ of such an ensemble $\rho$ can
be arbitrarily closely approximated by its (renormalised) first $M$
terms, corresponding to a discrete ensemble $\rho_M$ with a finite
index set. Hence, from from the assumed continuity of ensemble
volume and Eqs.~(\ref{less}) and (\ref{more}), $V(\rho) = V_I
\lim_{M\rightarrow\infty} \exp [S(\rho_M )]$ where $V_I$ is the
volume of a ``pure'' ensemble with respect to the infinite index set
$I$.  Thus $V(\rho)$ is as per the theorem (but becomes infinite in the
case that the limit of $S(\rho_M )$ as $M\rightarrow\infty$
does not exist).

The extension to quantum ensembles is straightforward.  Indeed, for
quantum ensembles the above analysis goes through formally unchanged,
where the expansion in
Eq.~(\ref{rhomix}) is now identified with an orthogonal decomposition
into pure states, and the first product in Eq.~(\ref{rhoalpha}) 
is a tensor product. Thus
the $\rho_i$ and $p_i$ represent (non-overlapping) eigenstates and
eigenvalues of $\rho$. The only additional consideration
is that $V_I$, the volume of an eigenstate of $\rho$, might
conceivably depend on the eigenstate basis.  However this is ruled
out by the invariance property (i): {\it all} pure states on a given
Hilbert space can be 
connected by unitary transformations, and hence have the same volume.
 
Finally, the theorem may be extended to continuous classical ensembles 
as follows.  Consider a classical ensemble $\rho$ described by a 
probability distribution $p({\bf x})$ on an 
$n$-dimensional space $X$. This
space may be partitioned into a set $\{ S_i\}$ of non-overlapping sets
of equal volume $V$ (i.e., $\int_{S_i} d^n{\bf x} = V$ 
for all $i$).  Define
the corresponding ``pure'' ensembles $\rho_i$ by the associated
probability distributions $p^{(i)}({\bf x}) = 1/V$ 
for ${\bf x}\in S_i$ and $= 0$ for ${\bf x}\notin S_i$.
These pure ensembles can be mapped to each other
by measure-preserving transformations, and hence from the invariance 
property have equal ensemble volumes, $V_0 (V)$ say.  The
formal analogues of the properties in Eq.~(\ref{pure}) then hold, and
again the above analysis for classical discrete ensembles goes 
through formally unchanged for mixtures of these pure ensembles, i.e.,
\begin{equation} \label{lem} 
V(\sum_i p_i \rho_i ) = V_0(V) \exp (-\sum_i p_i \log p_i ) . 
\end{equation}

Now consider the particular mixture defined by 
\begin{equation}
\rho_V = \sum_i p_i (V) \rho_i  , \hspace{1cm}
p_i (V) = \int_{S_i} d^{n} \, p({\bf x})  .
\end{equation}
Thus $\rho_V$ is a discrete approximation to $\rho$, and hence, noting
that $\int_X d^n{\bf x} \equiv \sum_i \int_{S_i} d^n{\bf x}$,
one has from the Mean Value Theorem that
\begin{equation} \label{mex}
{\rm Tr}_\Gamma [\mid \rho -\rho_V \mid ] = \sum_i \int_{S_i} d^n{\bf x}
\, \mid p({\bf x}) - p_i (V) /V \mid \rightarrow 0
\end{equation} 
in the continuum limit $V\rightarrow 0$. Hence, from Eq.~(\ref{lem}) and
the assumed continuity of  ensemble volume, 
\begin{equation} \label{tex}
V(\rho ) = \lim_{V\rightarrow 0} V_{0}(V) \exp (S_V )   
\end{equation}
where $S_V$ denotes the entropy of $\{ p_i (V)\}$.
But again approximating an integral by a summation, 
\begin{equation}
S(\rho) = \lim_{V\rightarrow 0}
- V \sum_i [p_i(V)/V] \ln [p_i(V)/V] \\
=\lim_{V\rightarrow 0} (S_V + \ln V ) . 
\end{equation}
Hence Eq.~(\ref{tex}) can be rewritten as
\begin{equation} \label{texx}
 V(\rho ) = e^{S(\rho)} \,\lim_{V\rightarrow 0} V_{0}(V)/V . 
\end{equation}
Finally, to show that the limit exists in Eq.~(\ref{texx}), note that 
any set $S\in X$ of measure $\int_S d^n{\bf x} = V$ 
can be partitioned into
$m$ non-overlapping sets of equal measure $V/m$ for any integer $m$.
Moreover, a ``pure'' ensemble on $S$, 
corresponding to a distribution which is uniform over $S$ 
and vanishing outside $S$, can trivially be written as an  
equally-weighted mixture of analogously defined ensembles for the
members of the partition.  Hence from the additivity property one has
the relation $V_{0}(V) = m V_{0}(V/m)$ for the ensemble volumes of
``pure'' ensembles. Further, replacing $V$ by $nV$ in this relation for
any integer $n$ implies that $V_{0}(rV) = r V_{0}(V)$ for any rational
number $r=n/m$.  This can be extended to all real $r$ from the assumed
continuity of ensemble volume, so that $V_{0}(V)/V =$ constant $=
K(\Gamma)$ say, and the theorem follows via Eq.~(\ref{texx}).

\newpage 

{\begin{table} {TABLE I. Examples of ensemble length 
and RMS deviation}\\
\renewcommand{\thefootnote}{\alph{footnote}}
\begin{tabular}{|llcc|}
\hline
Distribution & $p(x)$ & $L_{X}$ & $\Delta X $ \\
\hline
Uniform & $p_{U}(x)=1$,  $0\leq x\leq 1$ & $1$ & $1/(2\sqrt 3)$ \\
Circular & $p_{C}(x)=2(1-x^2)^{1/2}/\pi$,  $\mid x\mid \leq 1$ &
$\pi/\sqrt e$ & $1/2$ \\
Gaussian & $p_{G}(x)=(2\pi)^{-1/2}\exp(-x^2/2)$ & $(2\pi e)^{1/2}$ &
$1$ \\
Exponential & $p_{E}(x)=\exp (-x)$, $x\geq 0$ & $e$ & $1$ \\
Sink-squared & $p_{SS}(x)=\pi^{-1}[\sin (x)/x]^2$ & $\pi e^{2(1-C)}$
\footnotemark & -  \\
Cauchy-Lorentz & $p_{CL}(x)=\pi^{-1}/(1+x^2)$ & $4\pi$ & -  \\
Double-uniform & $p_{DU}(x)=1/2$,  $0\leq\mid x\mid - a\leq 1$ &
$2$ & $[1/3 + a(a+1)]^{1/2}$ \\ \hline
\end{tabular}\\
\addtocounter{footnote}{-1}
\footnotemark {$C\approx 0.57721566$ denotes Euler's constant}
\end{table}}\

\newpage

{\bf FIGURE CAPTIONS}
\\
\\
\\
FIG. 1.  Two uncorrelated ensembles $\rho_{1}$ and $\rho_{2}$ on
spaces $\Gamma_{1}$ and $\Gamma_{2}$ respectively (shown here
compressed to 1-dimensional axes), have respective volumes
$V(\rho_{1})$ and $V(\rho_{2})$ as indicated by the darkened axis 
regions.  The {\it Cartesian property} Eq.~(\ref{cart}) states that the
corresponding joint ensemble $\rho$ has a ``rectangular"
volume $V(\rho)=V(\rho_{1})V(\rho_{2})$, i.e., $V(\rho)$
corresponds to the Cartesian product of volumes $V(\rho_{1})$
and $V(\rho_{2})$.
\\
\\
\\
FIG. 2.  An ensemble $\rho$ on the product space of  $\Gamma_{1}$
and $\Gamma_{2}$ has a volume $V(\rho)$ indicated by the 
solid closed curve.  The corresponding projected 
ensembles  $\rho_{1}$ and $\rho_{2}$ on  $\Gamma_{1}$ and $\Gamma_{2}$
respectively have projected volumes  $V(\rho_{1})$ and
 $V(\rho_{2})$, indicated by the darkened axis regions.  The
{\it projection property} Eq.~(\ref{proj}) 
states that $V(\rho)$ can be no
greater than the volume of the rectangular region formed by the 
dashed lines, i.e., than the product of the projected
volumes.


\newpage
\begin{thebibliography}{99}

\bibitem[1]{eprint} M.J.W. Hall, quant-ph/9806013.

\bibitem[2]{shannon} C.E Shannon, Proc. IRE {\bf 37}, 160 (1949),
reprinted in {\it Claude Elwood Shannon: Collected Papers}, edited by
N. Sloane and A. Wyner (IEEE, New York, 1993), pp. 160-172.

\bibitem[3]{cd} C.M. Caves and P.D. Drummond, Rev. Mod. Phys.
{\bf 66}, 481 (1994).

\bibitem[4]{hallpra} M.J.W. Hall, Phys. Rev. A {\bf 55}, 100 (1997).

\bibitem[5]{statmech}  M. Toda, R.Kubo and
N. Sait\^o {\it Statistical Physics I} (Springer, Berlin, 1983),
Sec. 2.1.

\bibitem[6]{ma} S.-K. Ma, {\it Statistical Mechanics} (World Scientific,
Singapore, 1985), Secs. 5.2, 25, 26.

\bibitem[7]{leipnik} R. Leipnik, Inform. and Contr. {\bf 2}, 64 (1959).

\bibitem[8]{manko} V.V. Dodonov and V.I. Man'ko, in {\it Invariants and
the Evolution of Nonstationary Quantum Systems}, edited by M.A. Markov
(Nova Science Pubishers, New York, 1989), pp. 3-101, Sec. 4.
 
\bibitem[9]{zakai} M. Zakai, Inform. and Contr. {\bf 3}, 101 (1960).

\bibitem[10]{zed} K. Zyczkowski, J. Phys. A {\bf 23}, 4427 (1990).

\bibitem[11]{mk1} B. Mirbach and H.J. Korsch, Phys. Rev. Lett. {\bf 75},
362 (1995).

\bibitem[12]{mk2} B. Mirbach and H.J. Korsch, Ann. Phys. (NY) {\bf 265},
80 (1998).

\bibitem[13]{hell} E. Heller, Phys. Rev. A {\bf 35}, 1360 (1987).

\bibitem[14]{zurek} W.H. Zurek, S. Habib, and J.P. Paz, Phys. Rev. Lett.
{\bf 70}, 1187 (1993). 

\bibitem[15]{renyi} A. Renyi, {\it Probability Theory} (North-Holland,
Amsterdam, 1970), Theorem 4 of Sec. IX.6, Eq. (31) of Sec.IX.8.

\bibitem[16]{thesis} J.B.M. Uffink {\it Measures of Uncertainty and the
Uncertainty Principle} (Ph.D. thesis, University of Utrecht, 1990).

\bibitem[17]{maas} H. Maassen and J.B.M. Uffink, Phys. Rev. Lett.
{\bf 60}, 1103 (1988).

\bibitem[18]{foot1} To derive Eq.~(\ref{length}), define $f(\alpha)$ $=$
$\ln \int dx \, p(x)^{1+\alpha}$, implying $\lim_{\alpha\rightarrow 0}$
$\ln L_{X,\alpha}$ = $\lim_{\alpha\rightarrow 0}$ $[f(0)-f(\alpha )]
/\alpha$ = $-f'(0)$.

\bibitem[19]{ash} R. Ash {\it Information Theory} (Wiley, New York, 
1965),  Chaps. 1,2,6.

\bibitem[20]{foot2} This follows from a straightforward variational
calculation, and may be generalised to all Renyi lengths.
  
\bibitem[21]{foot3}  For the general case of non-overlapping $p(x)$ and 
$q(x)$ with equal Renyi lengths $L_{X,\alpha}=L$, one finds for the
mixture $\lambda p(x) + (1-\lambda )q(x)$ that  $L_{X,\alpha}/L$ $=$
$[\lambda^{1+\alpha} + (1-\lambda)^{1+\alpha}]^{-1/\alpha}$;
differentiating with respect to $\lambda$ yields a maximum of 2,
attained for $\lambda=1/2$.

\bibitem[22]{foot4} For an example in the context of optical phase, 
where the RMS deviation gives a rather misleading indication of the 
spread of the (near-optimally phase-concentrated) coherent phase states,
see M.J.W. Hall, J. Mod. Opt. {\bf 40}, 809 (1993).

\bibitem[23]{bbm} I. Bialynicki-Birula and J. Mycielski, Commun. Math.
Phys. {\bf 44}, 129 (1975).

\bibitem[24]{hardy} G.H. Hardy, J.E. Littlewood, and G. Polya,
{\it Inequalites} (Cambridge University Press, London, 1934), Eq.
(2.13.7).

\bibitem[25]{footrms} If the components of ${\bf x}$ are (possibly
non-commuting) quantum observables, the expectation values in
Eq.~(\ref{rmsarea}) should be symmetrised.

\bibitem[26]{shan} C.E. Shannon, Bell Syst. Tech J. {\bf 27}, 379
(1948); {\bf 27}, 623 (also reprinted in the collection of Ref. [2] 
above,  pp. 5-83).

\bibitem[27]{foot5} Of course, in special contexts, restricted 
interpretations of ensemble  
entropy are possible: for example, as the logarithm of the number
of microstates in statistical
mechanics \cite{statmech}; or as the amount of information 
obtainable by measurement on discrete classical and quantum
ensembles \cite{shan,tight} (see also
L. Brillouin, {\it Science and Information Theory} 
2nd edn. (Academic, New York, 1962), Chapter 1).
Such interpretations are not applicable more generally, however.  For
example, for continuous classical ensembles the entropy can take
any negative value, which is not the case for numbers of microstates
and information gain. In contrast, in the geometric interpretation
such negative values merely correspond to the possibility of 
arbitrarily small ensemble volumes.

\bibitem[28]{footbit} Units of bits correspond to dividing $S(\rho)$ in
Eq.~(\ref{ent}) by $\ln 2$.

\bibitem[29]{behara} M. Behara, {\it Additive and Nonadditive Measures 
of Entropy} (Wiley, New Delhi, 1990), Section 1.1.

\bibitem[30]{relent} M. Ohya and and D. Petz, {\it Quantum Entropy and 
its Use} (Springer-Verlag, Berlin, 1993).

\bibitem[31]{plenio} V. Vedral and M.B. Plenio, Phys. Rev. A {\bf 57},
1619 (1998), Sec. IV.

\bibitem[32]{semiclass} A.S. Davydov, {\it Quantum Mechanics} 2nd edn.
(Pergamon Press, UK, 1976), Sec. III.23.

\bibitem[33]{footsig} The signals are in general represented 
by ensembles, to take into
account any noise processes in the transmitter and channel medium prior
to detection at the receiver.

\bibitem[34]{holevo} A.S. Holevo, Probl.
Inf. Trans. {\bf 9}, 177 (1973).

\bibitem[35]{others} H.P. Yuen and M. Ozawa, Phys. Rev. Lett. {\bf 70},
363 (1993); C.A. Fuchs and C.M. Caves, Phys. Rev. Lett. {\bf 73},
3047 (1994); B. Schumacher, M. Westmoreland, and W.K. Wootters, Phys.
Rev. Lett. {\bf 76}, 3452 (1996).

\bibitem[36]{tight} B. Schumacher and M.D. Westmoreland, Phys. Rev. A
{\bf 56}, 131 (1997);  A.S. Holevo, quant-ph/9611023.

\bibitem[37]{husimi} A complete set of coherent states $\{\mid\alpha
\rangle\}$ corresponds to a joint position/momentum measurement 
described by the positive-operator-valued measure $\{\pi^{-1}
\mid\alpha\rangle\langle\alpha\mid\}$ (see, e.g., \cite{cd}); the
statistics of the corresponding observable $A_Q$ for state $\rho$ are
then given by the Husimi distribution $\pi^{-1}\langle\alpha\mid
\rho\mid\alpha\rangle$.

\bibitem[38]{hallprl} M.J.W. Hall, Phys. Rev. Lett. {\bf 74}, 3307
(1995).

\end{thebibliography}
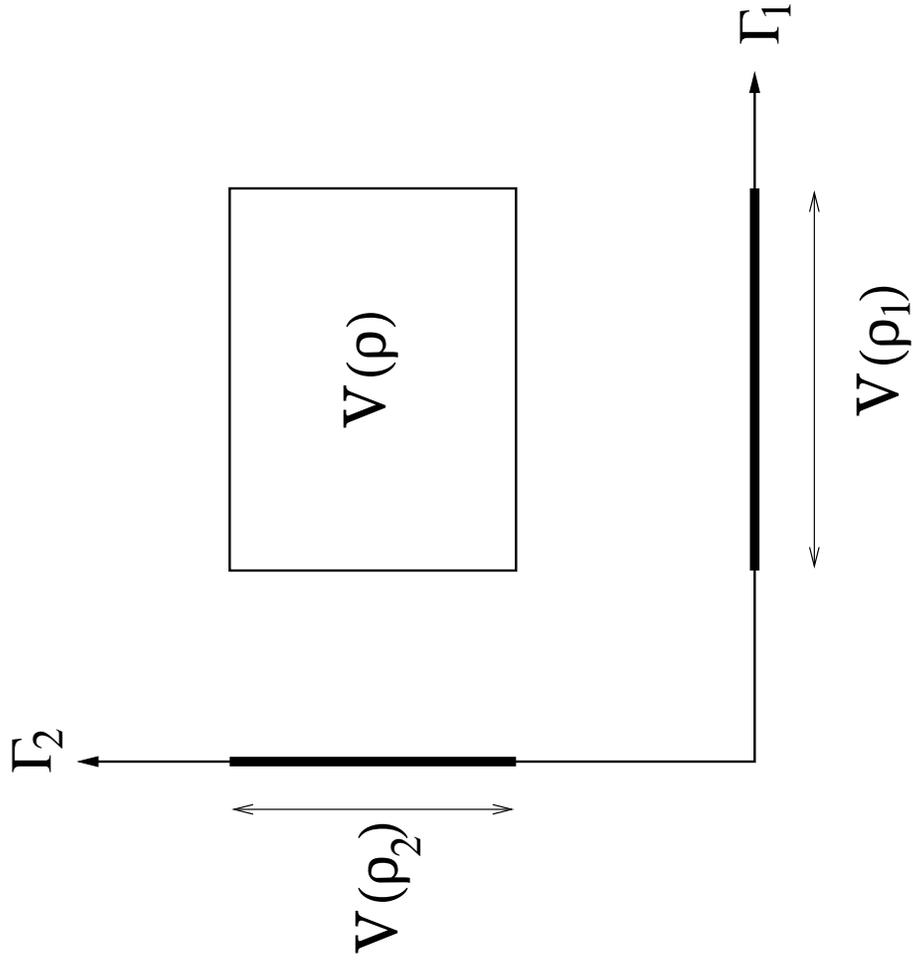
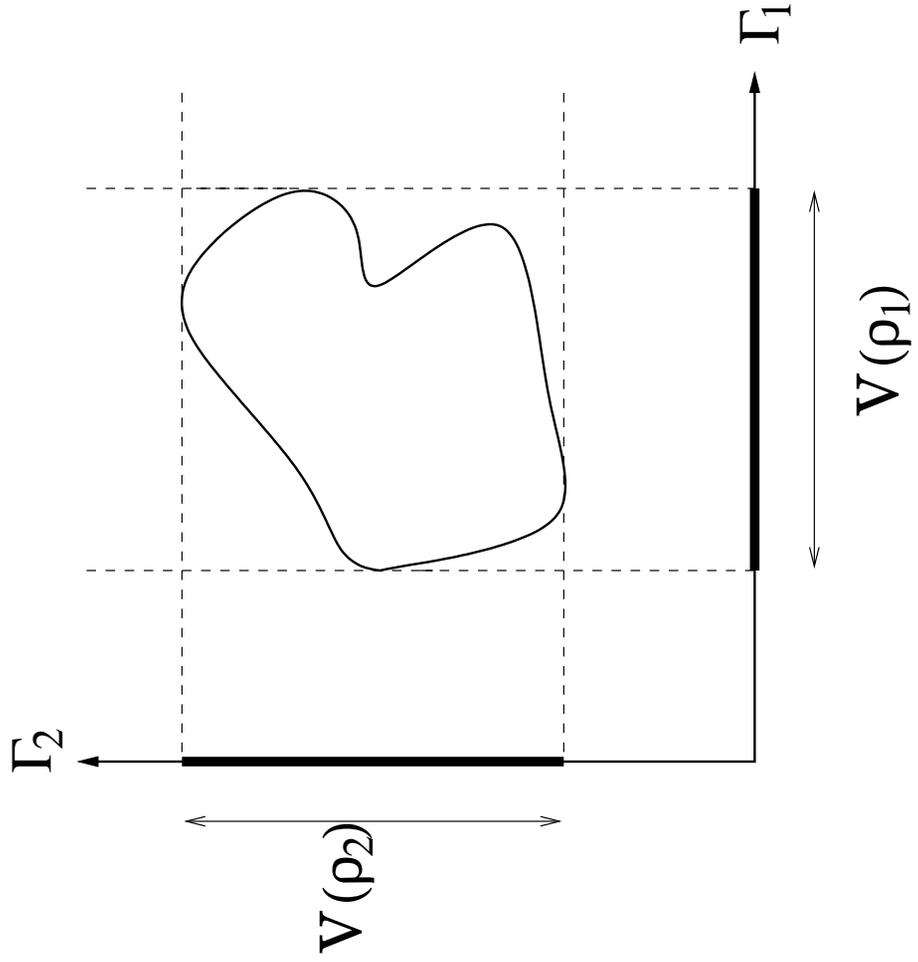
\end{document}